\documentclass[conference]{IEEEtran}
\ifCLASSINFOpdf
\else
\fi
\begin{document}
\title{Attacking Machine Learning Models \\as Part of a Cyber Kill Chain}

\author{
\IEEEauthorblockN{Tam N. Nguyen}
\IEEEauthorblockA{North Carolina State University\\
tam.nguyen@ncsu.edu\\}
}

\maketitle
\begin{abstract}
Machine learning is gaining popularity in the network security domain as many more network-enabled devices get connected, as malicious activities become stealthier, and as new technologies like Software Defined Networking emerge. Compromising machine learning model is a desirable goal. In fact, spammers have been quite successful getting through machine learning enabled spam filters for years. While previous works have been done on adversarial machine learning, none has been considered within a defense-in-depth environment, in which correct classification alone may not be good enough. For the first time, this paper proposes a cyber kill-chain for attacking machine learning models together with a proof of concept. The intention is to provide a high level attack model that inspire more secure processes in research/design/implementation of machine learning based security solutions.
\end{abstract}


\IEEEpeerreviewmaketitle

\section{Introduction}
There is a significant gap between the amounts of connected devices and the number of cyber security professionals. Per U.S. Bureau of Labor Statistics \cite{USBureauofLaborStatistics2016}, a projected growth in cyber security jobs from 2014 to 2024 is 18\% while Cisco \cite{Forecast2017} predicted a 100\% increase in network-enabled devices, growing from 4 billions in 2016 to 8 billions units in 2021. Consequently, global data traffics will increase by at least 5 times. Furthermore, the emergence of Software Defined Network makes it very convenient to develop and deploy machine learning (ML) based network security solutions, some of which are shown in Table \ref{table:MLsolutions}. Usually, researchers design new ML-based security solutions and then use benchmark results to prove the models' accuracy. Such methodology alone is not attractive enough in the eyes of cybersecurity leaders due to the lack of information on how those solutions will fit into the bigger picture at their organizations. Other than accuracy, the cybersecurity leaders would also like to know about the model's projected maintenance costs, the model's ability to withstand abuses, the quality of source codes, the ways datasets were collected, and so on. For that reason, this paper offers several recommendations on how to improve the overall practical robustness of a machine learning enabled security solution. The main contribution is the ML cyber kill-chain - a high level threat model tailored specifically for ML. The organization of this paper follows the Russian doll approach. It starts with the biggest picture - the Security Operation Center (S.O.C) with their missions, operations and problems. It is then followed by a brief survey on the inherent limitations of existing ML algorithms being used by S.O.C. Within that sub-picture, the paper formalizes ML specific threats into an attack model - the ML cyber kill chain. Finally, the paper proposes a list of recommendations for a more secure process of designing new ML-based security solutions.


\hfill tnn

\hfill April 05, 2016

\begin{table*}[!t]
  \centering
  \begin{tabular}{ p{5cm}  p{2.2cm}  p{1cm}  p{8.1cm} }
    \hline
    \\
    PUBLICATION & FUNCTIONALITY & SUCCESS & ML METHODOLOGY \\
    \\
    \hline
    \\
    AMPS: Application aware multipath flow routing using machine learning in SDN  \cite{Pasca2017AMPS:SDN} & Dynamic multipath flow routing & 98\% & Supervised learning on 40-features dataset with C4.5 Decision Tree algorithm \\
	\\
    Dynamic attack detection and mitigation in IoT using SDN \cite{Bhunia2017DynamicSDN} & IoT monitoring and protection & 98\% & Support Vector Machine model was used on processed data passed down from a learning module in order to classify traffics. Both linear and non-linear kernel (RBF) were used.\\
    \\
    Analytics-Enhanced Automated Code Verification
for Dependability of Software-Defined Networks \cite{JagadeesanAnalytics-EnhancedNetworks} & Detect and prevent malicious SDN app behaviors & 99\% & It is a combination of automated code verification (Java Path Finder) with ML analytic model (Ckmeans), being independent of underlying network topology\\
    \\
    An applied pattern-driven corpus to predictive analytics in mitigating SQL injection attack & Prevent SQL injection in the cloud & 98\% & Two-Class Support Vector Machine and Two-Class Logistic Regression with linear kernel implemented on Microsoft Azure Machine Learning \\
    \\
    Forecasting and anticipating SLO breaches in programmable networks \cite{Bendriss2017ForecastingNetworks} & Cognitive SLA enforcement & 90\% & Protecting Service Level Objectives (availability, response time, throughput) using Long Short Term Memory Recurrent Neural Network (able to identify previously seen patterns in new distorted samples) \\
    \\
    Access point selection algorithm for providing optimal AP in SDN-based wireless network \cite{Lee2017AccessNetwork} & Pick best QoS, less backhaul conges- tion  & depends & Traffic classification deployed at AP, using C5.0 Decision Tree algorithm \\
    \\
    Content Popularity Prediction and Caching for ICN: A Deep Learning Approach With SDN  \cite{Liu2018ContentSDN} & Cache management & 75\% mean accuracy & Improving cach operations by predicting the popularity of content using Deep-Learning-based Content Popularity Prediction (Stacked Auto-Encoders + Softmax)  \\
    \\
    AWESoME: Big Data for Automatic Web Service Management in SDN \cite{Trevisan2018AWESoME:SDN} & Web traffic engineering & 90\% & Annotation module at network edge classifies flows in real time based on bag-of-domains (using Apache Spark), flow-to-domain, domain-2-service, and service-to-rule. \\
    \\
    Athena: A Framework for Scalable Anomaly Detection in Software-Defined Networks \cite{Lee2017Athena:Networks} & Network security development framework & depends & 11 ML models were made available as library, allowing developers to quickly develop network security applications that can perform real-time detection and responses. \\
    \\
   FADM: DDoS Flooding Attack Detection and Mitigation System in Software-Defined Networking \cite{Hu2017FADM:Networking} & framework for DOS prevention & depends & Feature extraction is entropy-based and attack detection is powered by Support Vector Machine  \\
    \\
    \hline \\
  \end{tabular}
  \caption{10 ML-based security solutions for SDNs since 2017}
  \label{table:MLsolutions}
\end{table*}

\section{Backgrounds on S.O.C Processes}
Security Operation Center (S.O.C) is part of a "Defense in depth" strategy. Metaphorically, "defense in depth" is like an artichoke, consisting of interlaced, overlapping-but-independent protection layers backing each other. When some of its layers got pealed away, an artichoke still maintain almost the same shape (posture). In response, adversaries employs "advanced persistent" attack strategies in which persistent organized efforts can be categorized into phases also known as "intrusion kill chain" \cite{Hutchins}. 
\subsection{The General Cyber Intrusion Kill Chain}
\begin{enumerate}
\item Reconnaissance (recon) : adversaries do research about the targeted system's structures, capabilities, vulnerabilities, etc
\item Weaponization : adversaries code up deliverable payloads
\item Delivery : adversaries use different ways to deliver weaponized payloads to the destination
\item Exploitation : payloads were executed and vulnerabilities of target system(s) were exploited
\item Installation : malware(s) were installed and the "persistent" factor can now be established
\item Command and Control (CnC) : adversaries established a hidden control channel with compromised entities within the victim system to further organize and expand the attack campaign
\item Actions : adversaries now perform actions on their true objectives
\end{enumerate}
S.O.C daily operations align very closely with cyber kill chains for several good reasons. It reminds S.O.C analysts about the big picture in which locating an attack should be looked at as part of a past-present-future orchestrated campaign. For example, if an installation of a malware was identified (phase 5 of the kill chain), it is crucial to traverse to earlier phases and find out: what vulnerability(ies) in what subsystem was exploited? In what way the initial attack payload was delivered? How the payload was built? and how did the attackers know of such vulnerability(ies) ? In the mean time, analysts also need to look forward and try to predict how attackers will use the piece of malware to further exploit the system. There are usually patterns within each phase of the kill chain, and between kill chains of different attack campaigns. An advanced persistent attack campaign is usually a combination of those patterns \cite{Hutchins}. Therefore, in-depth knowledge of past kill chains will help with detecting, preventing and predicting future kill chains. Since all of those heavy analytical tasks cannot be carried out by humans alone, Machine Learning (ML) models are used. Most of ML related work-flows in S.O.C can be categorized into 3 common steps : Data pre-processing, clustering/threshold adjustment, and monitoring.

\subsection{Data pre-processing}
Raw captured data is huge and came from various sources such as full packet capture (PCAP), NetFlow (generated by Cisco network devices), protocol metadata, application logs, machine logs, telemetry streams. Just within a NetFlow record itself, there are 104 possible data fields with 22 additional fields reserved by Cisco\cite{Cisco2011NetFlowServices}. Common methods are dimensional reduction (mapping more dimension variables into fewer ones), clustering (identifying groups of items with similar characteristics), Spearman's rank correlation \cite{Blowers2014}, statistical sampling, measuring and pick samples based on entropy and so on. \cite{SultanaSurveyApproaches}. Common issues include but are not limited to streaming fact issue, noisy data, and the trustworthiness of data. Streaming fact issue refers to wrong data clustering decision based on incomplete facts (key information has not yet arrived). Some systems try to overcome this problem by using ML models to make inferences as data are coming in. Noisy data and untrustworthy data can be caused by hackers or by aging devices, network congestions, or simply, configuration mistakes made by field admins.
\subsection{Clustering/threshold adjustment}
It is estimated that the amount of false positive alarms raised by signal based IDS is 5 to 20 times the amount of true positive ones \cite{Blowers2014}. Having too many false positive alarms will cost time, money and corrode the credibility of the intrusion detection system. Therefore, methods like Adaptive Cluster Tuning process (ACT) is used by S.O.C to adjust the initial categorization/clustering done by machines \cite{Blowers2014}.
\subsection{Monitoring}
Finally, models are deployed. Traffics are monitored and packages got sent to first stage for pre-processing - the cycle repeats. It is important to do so in order to identify concept-drift due to evolution of attacking methods or to identify new patterns and concept-evolution.
\newline \newline
Besides the above-mentioned issues, S.O.C also has to balance the goal of early attack detection and the cost of false positive alarms; to catch up with the speed of evolving attacks while tuning the existing system. While the human factor can help the ML models in many ways, it is also important to note that human is another attack surface. Human makes mistakes.

\section{Common Machine Learning models \\Used in S.O.C}
In daily operations, S.O.C should train and re-train several models simultaneously. It is about the spirit of "defense-in-depth" since no one model is perfect for identifying every security problems as illustrated in the work of Buczak \cite{Buczak2016}. While some statistics gathered by Buczak will be presented for the convenience of illustrating the general effectiveness of each model, the best way to really evaluate and compare ML models is testing them in one's own lab environment with the same analysts tunning the models.

\subsection{Artificial Neural Networks}
By design, ANNs are fit for non-linear problems but tend to suffer from local minima leading to long learning time, and as the number of features increases, the longer it will take to learn. With anomaly and hybrid detection, multi layer ANNs were used to analyze data. The models can reach deeper into lower network layer data and are able to detect some low/slow type attacks. However, the performance of ANNs is not consistent. While they can identify 100\% of the normal behavior, the amount of false alarms may sometimes reach 76\% depending on what kind of attacks were being executed \cite{Buczak2016}.

\subsection{Bayesian network}
Bayesian network is a probabilistic directed acyclic graph type with nodes as variables and the edges as their relationships. Based on the relationships, a node can "walk" to another. Each node has a probabilistic value and at the end of the walk, a final probabilistic score is formed. Relationship links that have high true positive score will be verified and formed into rules. Therefore, a Bayesian network is proactive even in misuse mode. In a test of using model to label IRC-botnet generated data, the precision rate is 93\% with a false positive rate of 1.39\% (detecting fewer cases than some other models but generating less false alerts). In other tests, the reported precision rates are 89\%, 99\%, 21\%, 7\%, and 66\% for DoS, probe/scan, remote-to-local, user-to-root, and "other" classes of attacks respectively \cite{Buczak2016}. In the case of anomaly detection, ACT process was used to tune the system. Accuracy of 100\% with 0.1\% false alarm rate were reported in lab experiments analyzing TCP/IP packets.

\subsection{Clustering}
Some popular clustering models are k-means, k-nearest neighbor, density-based spatial clustering of applications with noise (DBSCAN), etc. Because the models were designed in order to find patterns in unlabeled multi-dimensional data, explicit descriptions of classes are not required. A weakness of this model is known as the "curse of dimensionality". Too many features may confuse the model and any imbalance in the feature set will negatively affect its decisions. In some experiments, this method can detect up to 80\% of unknown attacks. In anomaly mode, studies suggest that clustering models can be really accurate (98\%) in analyzing captured PCAP packages but performance goes down when dealing with streaming data (false alarm rate may go up to 28\%)

\subsection{Decision trees}
Decision tree is a flow-chart like structure built on concepts of information gain/entropy where each node choose the best fit attribute to split current set of examples into subsets. Normally, decision trees provide the benefits of high accuracy with simple implementation. However, it is not usually the case with larger trees. Also with large, complex trees, the model tends to favor attributes with more levels. To overcome issues with large trees, analysts will have to do some pruning to get smaller trees. In anomaly detection mode, experimental results of using decision tree model to detect bad domain names from passive DNS queries show that the model is accurate with acceptable false alarm rates (reduced by constant re-training).

\subsection{Evolutionary computation}
While Genetic Algorithm (GA) and Genetic Programming (GP) are most used Evolutionary Computation (EC) methods; Particle swarm optimization, Ant Colony Optimization, Evolution Strategies are also parts of the group. The main concept is based on the idea of "the strongest will prevail" and basic operators are selection, crossover, and mutation. Experiments with various attack types show that the average false alarm rate is very low. However, the sensitivity in detecting new attacks varies greatly (from 66\% to 100\%) depending on attack types.

\subsection{Naive Bayes}
Naive Bayes model calculates the final conditional probability of "attack" (or "normal") with a naive assumption that the used features are independent from each other. That assumption is the biggest limitation of this model. However, if the features are indeed independent from each other, naive bayes can be very powerful thanks to its simple algorithm that allows the model to be highly scalable and be used as an online classifier. In misuse mode, experimental results suggest a decent accuracy (above 90\%) but with a quite unacceptable false alarm rate (around 3\%). In anomaly detection mode, experimental results show a tremendous difference in accuracy. For example, accuracy of identifying data as "normal" is 97\% for DoS type attacks and only 9\% for remote-to-local type attacks. 
 
\subsection{Supported vector machine}
Supported Vector Machine (SVM) is a binary classification model by design. With a kernel such as linear, polynomial, Gaussian Radial Basis Function, or hyperbolic tangen; the model will try to draw a hyperplane that divides the feature space into two classes. Sometimes, when overlapping is unavoidable, slack variables will be added and each overlapping data point will be be assigned a cost value. In misuse detection experiments, a large set of features is reduced by using feature selection policies or feature selection algorithms. The model is quite accurate but also shows limitations at identifying certain types of attacks such as user-to-root attack. In anomaly detection mode, usually SVMs will use more sophisticated kernel to help with the drawing of the hyperplane. Experimental results show great variations in accuracy (from 65\% to 99.9\%) and sometimes, false negative rate can get really high (over 30\%) \cite{Buczak2016}.

\section{Attacks on \\Machine Learning  models}
The definition of "attack" on ML models should be flexible and be focusing on the models' purposes rather than the models' functionalities. The accuracy rate is not the only thing adversaries can target. In one case, if a model was designed to allow or drop suspicious packets then reducing the model's accuracy fits the definition of an attack. In another case, adversaries can cause the models to produce true positives that are very close to false positives. Consequently, it causes burn-outs on the security analysts who are going to manually inspect those flags. The model's functionalities are intact but one of its purposes - reducing the security analysts' workloads - was compromised. "Attack" can also mean significantly increasing the time it takes for a ML model to make a specific decision. No matter what attack end-goals are, attackers must first have some really good insights about the targeted model. Thus, there is a strong motivation to clone/extract a security ML model.

When performing extraction, inputs are given to a trained model, end results (outputs) got harvested, and clone model learns from those input-output data pairs. While it appears that training the original model and cloning an existing model are quite similar, model cloning does not have to deal with broken or faulty data entries that delay or even mislead the learning process. Model cloning also does not have to deal with optimization issues such as local minima/maxima traps. In 2016, Tramer et. al. \cite{Tramer2016} proposed several methods to perform model extraction of several ML types but those methods were not weaponized. For example, it required Tramer and his team to perform thousands of probes in order to extract a model and such number of probe is too high to be considered practical in targeting a protected environment. In the following sections, we will briefly discuss Tramer's methods and then propose a way to weaponize them.

\subsection{Equation-solving attack}
This form of attack is fit for logistic regression types such as binary logistic regression (BLR), multi-class logistic regression (MLR), and multi-layer perceptron (MLP). Because the models can be represented as equations with variables, attackers just need to feed the known variable values in, use mathematics to solve the equations and get the rest of the unknown values. For example, with BLR, we have : \newline \quad\quad \(w \in R^{d}, \beta \in R\) with \(f_{i}(x) = \sigma(w \times x + \beta) \) \newline where \newline
\(\sigma(t) = 1/(1+e^{-t})\) \newline
Attacker will feed \(x_{i}\) to the trained model and the model will give \(y_{i} = f(x_{i}) = \sigma(w \times x_{i} + \beta) \). If we have enough \({x_{i},y_{i}}\), we should be able to solve the equations to get \(w\) and \(\beta\). The math will be more complicated when dealing with MLRs and MLPs.With softmax model in MLR for instance, we have:\newline
\(c>2, w \in R^{cd}, \beta \in R^{c}\)\newline
\(f_{i}(x_{i}) = e^{w_{i} \times x + \beta_{i}}/(\sum_{j=0}^{c-1}e^{w_{j} \times x + \beta_{j}})\) \newline
The function can be solved by minimizing its cost/loss functions. The methods proved to be effective in Tramer's experiments \cite{Tramer2016}. With BLR, they were able to achieve \(R_{test}=R_{unif}=0\) with an average probe of 41. The number of probe is much greater with MLR and MLP. For instance,it required them to perform 54,100 queries on average in order to achieve 100\% accuracy on a 20 hidden node MLP. Unlike BLR, it is sometimes very hard to estimate a correct amount of probes needed for MLP and MLR cloning. Especially with MLP, it is hard for attackers to guess how many hidden neuron layers are there, and how many neurons per layer. Attackers will also not be able to tell how many classes an original MLP can identify. However, everything can be different in actual cyber attack scenario. Instead of 100\% accuracy, attackers may only need to clone a model with 90\% accuracy for their purposes, and the amount of probes needed may be significantly lower. Another reason to not aiming for 100\% accuracy is that original ML models get tunned on a fast pace, daily. A 100\% accurate cloned model of today maybe different from the actual model next week.
\subsection{Model inversion attack}
Given feature dimension \(d\) with feature vector \(x_{1},x_{2},...,x_{d}\), some knowledge about some of the features, and access to \(f\) - the model, Fredrikson et al. \cite{FredriksonModelCountermeasures} proposed that a black box model inversion attack which involves finding an optimal x that maximizes the probability of some known values.\newline
\(x_{opt} = argmax_{x \in X}f_{i}(x)\) \newline
For instance, if an image of Bob was used to train model M to recognize category "man". If that exact image is fed into M, the result will be "man" with 100\% confidence while images do not belong to the training set will never get such absolute score from the model. An attacker can start with one pixel and find the pixel value that gives the maximum score possible of category "man". The process continues to other pixels and the end result is an image very close to Bob's original image in the training set. Tramer et al. \cite{Tramer2016} upgraded this approach by performing inversion attack on a cloned model M' of M. The reported improvement is a 6-hour faster recovering time for 40 faces. This kind of attack opens a theoretical possibility of which attackers can gain some insightful knowledge about a security model's trained data set if they could clone the model with 100\% accuracy and somehow was able to tunnel it out. 
\subsection{Path-finding attack}
Tramer et al. \cite{Tramer2016} also extended prior works on tree attacks and proposed their "path-finding" attack which can be used to map binary trees, multi-nary trees, and regression trees. We have a tree \(T\) with \(v\) nodes and at each node, there is an identifier \(id_{v}\). With \(x \in X \), an oracle query will give \(O(x) = id_{v}\). If \(x \in X_{1} \cup {\perp} \times ... \times X_{d} \cup {\perp}\), \(O(x)\) will return the identifier at the node where T stops. To begin the attack, we pick \(x \in X_{1} \cup {\perp} \times X_{2} \cup {\perp} \times... X_{i1} = [a,b] \times ... \times X_{d} \cup {\perp}\). \(O(x)\) gives \(id_{Lv}\) at the leaves of the tree. We then can separate \([a,b]\) into n sub ranges where n is the number of the corresponding known leaves (at this point). For each \(X_{i1}\) sub range, we repeat the process and find another nodes/leaves. This was referred to as the top-down approach which is of higher performance than the bottom-up approach. Reported performance evaluations of this approach show that in order to achieve 100\% on \(1-R_{test}\) and \(1-R_{unif}\), it will take 29,609 queries to clone a tree with 318 leaves, 8 layers of depth; 1,788 queries to clone a tree with 49 leaves, 11 layers of depth; and 7,390 queries to clone a tree with 155 leaves and 9 layers of depth.

\subsection{Other attacks}
There are several other methods to attack ML models. The Lowd-Meek attack \cite{LowdAdversarialLearning} targets linear classifiers that give only class labels as models' outputs. The general idea of this approach is using adaptive queries to throw sample points at the suspected positions of the hyperplane. Another way to attack was described by Bruckner \cite{Bruckner} as a single Stackelberg Prediction Game (SPG). In this game, the Leader (L) is the one with the original ML model M. The Follower (F) is the attacker. F will attack L by generating and feeding model M data that at least will prevent M from learning new knowledge or at most, teach M new faulty knowledge. Theoretically, this can be achieved by providing learning data that maximize the cost function of model M. In real life situations, there are more than one attacker with different attacking goals and L does not know how many F are there and what exactly each F is trying to do. This escalates to the Bayesian Stackelberg Game. Zhou and Kantarcioglu described it as "Nested Stackelberg Game" \cite{ZhouModelingGames} suggesting a solution of using and switching a set of models to confuse the attacker. Kantarcioglu later on also proposed the concept of "planning many steps ahead" in this game. Details of these methods will be further studied and evaluated within a defense-in-depth environment and will be discussed in future works.

\section{The Machine Learning model \\kill chain}

Based on all the above-mentioned ML vulnerabilities and limitations, this paper proposes the following high level kill chain tailored specifically for ML based solutions. It is extremely important for the ML solution designers to be fully aware of this threat model and pay full attention to the little details that may get their solutions to be compromised. At S.O.C level, this threat model will help analysts to identify how a particular set of attacks fit into a bigger picture, how to properly document and categorize current/past attacks, and how to anticipate the attackers' next moves.

\begin{enumerate}
\item Recon phase\newline
The goal of this early phase is to gather as much basic information about a targeted ML system as possible. Important questions to be answered include "What ML model(s) being used?", "Are the models protecting against the kind of attacks I am planning to perform?" and "Is the targeted system based on an opensource project?" In reality, defenders will have certain favorite ML models in defending against certain types of attacks. We also know that open sourcing is the trend and most commercial cyber security systems have their own open source siblings. In such case, studying the anatomy of the open source system(s) may provide tremendous knowledge about the actual defense system being used. Probing efforts should be fragmented into small, seemingly uncorrelated attempts to avoid early detections. Social engineering or insider leak can be extremely helpful. Non technical information such as what model types are being used?, how many analysts within a certain shift? how many generated alerts created per day on an average? etc can be safely communicated-out without bursting the cover of the insider. 
\item Weaponization \newline
Based on collected basic knowledge about the targeted model, attackers at this stage will work on an optimized set of probes with the help of an adaptive engine. For example, if the model uses SVM, the values of malicious probes should be close to where the polynomial hyperplane goes. Because a model has to deal with many attackers at the same time, the actual polynomial hyperplane at the point of attack may have a shape that is slightly different from days before. In this situation, the adaptive engine will kick in, adjust the probes on the fly, and try to minimize the number of probes to a safe threshold.
\item Delivery \newline
Once attackers are confident that they have prepared the best set of data probes and a good adaptive engine to handle real-time data changes, they will launch the first wave of attacks. For example, the attack can target the decision border where the difference between true-positives and true-negative is  low. After a while, analysts may "tune" the model and relax the border and/or the model's cost functions resulting in some true-posstive to be accepted as false-positives. 
\item Exploitation \newline
Attackers now want to gather deeper data about the model and may as well expand probes to other models. It is reasonable to believe that at this point, attackers can deliver a full set of probes to clone the model and tunnel the data out thanks to the relaxed boundaries. Once having a functional clone in their own lab, attackers can perform full model inversion attacks in order to extract further details about the model and its trained data. Finally, a prediction model can be built from which, attackers can simulate certain attack scenarios with answers to: how the targeted system will react; how the model will be drifted; the rankings of threats etc.
\item Installation \newline
At this phase, attackers will use coordinated probes with the helps of the adaptive engine together with the prediction engine to poison the ML  model. The goal now is mis-training the targeted model and allow future attacks to happen.
\item Command and Control \newline
Now as the attackers have established a safe path to get through the ML Model, they will move on with setting up a hidden command and control channel with the helps of compromised entities within the victim's system to further organize and expand the attack campaign. This can be combined with the regular attack kill chain featuring malwares,botnet, etc.
\item Action\newline
Finally, attackers act on their main objectives.
\end{enumerate}

\subsection{Proof of Concept}
Let us take a look at an attack on a IBM cognitive classifier between domestic cats (simulating true negatives)  and female lions (simulating true positives). Choosing this type of dataset gives us several benefits. First, it is non-linear and is complicated enough, yet classification results are easy to be observed and understood. Second, there are many variations in between two main classes such as golden cat, jungle cat, cheetah, bobcat, lynx, etc which can be very helpful to simulate real life possibilities of attacks which may contain largely same features as legitimate instances and with only a few "harmful features". 
\begin{enumerate}
\item Setting up \newline
We will use IBM Watson Visual Recognition demo \cite{IBMIBMDemo} to train a model M. The Visual Recognition demo is a web interface maintained by IBM explicitly for demonstration purposes and is under IBM's control. Four models of M were trained using different sets of 10, 20, 50 and 70 samples per class for a little experiment within the next step. However, for the rest of this demonstration, M will be referred to as the model M that was trained with 50 samples. The name of each corresponding set of samples begins with "A1-1", "A1-2", "A1-5", "A1-7". Images used and recorded screen-capture videos can be found at \cite{TamNguyen2017ProjectEye}
\item Probe 1 \newline
At this point, we have the following scenario. A trained model that can detect normal traffics (cats) and attacks (female lions). An attacker - Bob - who wants to attack the model and he wants to start up carefully. Bob will use set A2 with different images of cats (6 images) and female lions (6 images) to probe M. Because Bob cannot afford a large amount of probes (at least not with lions), ideally Bob will have to carefully choose the samples in such a way that the samples help Bob explore the domains as much as possible. For example, with 6 attempts for cats, Bob should choose images such that they will be all classified as "cat" with different confidence scores of : "close to 50\%", "somewhere between 60\% and 80\%", and "higher than 80\%". However, in our experiment, we will choose probe pictures randomly. With each probe, M returns a classification  (A2-CL) and a confidence score (A2-CF). As previously mentioned, we run the probes on different versions of M, and as the number of training samples increases, there is not much improvement. For example, M was able to identify cat pictures with confidence score above 50\% with just 10 training samples for each category. However, when training samples gets to 70 for each category, confidence scores do not improve greatly and even fails to categorize picture named "c3". Future works will involve measuring the learning rates of this model using a better set of up to 1000 cat pictures.
\item Build a model predicting the other model
Based on pairs of A2-CL and A2-CF, Bob will set up his own model M' which, in this case, is also based on IBM Visual Recognition MLaaS. While the underlying technology is the same, M' is independent and different from M. In M', images in set A2 were uploaded with their associated A2-CL and A2-CF harvested from M. Images of cats and female lions are put into one "collection". The process is demonstrated in the second video at my project page.
\item Probe 2 \newline
At this point, Bob prepares his set of possible ways to attack the system. In between a normal use (cat) and an obvious attack (female lion), Bob came up with variations such as "bobcat", "cheetah", "jaguar", "tiger" etc (pictures in set A3). In his mind, Bob thinks "bobcat" maybe the best option to start with the first attack since it is very close to "cat". However, Bob decides to run his attack scenarios (pictures in set A3) through M' first. M' will compare the image (A3\(_{i}\)) with its image library and return the closest matched image (A2\(_{i}\)) with its A2\(_{i}\)-CL, A2\(_{i}\)-CF. M' will also return the score of how confidence M' thinks A3\(_{i}\) looks like A2\(_{i}\) - the A3\(_{i}\)A2\(_{i}\)-CF score. For each image (i) in A3 set, we calculate the total relational confidence score A3\(_{i}\)-CF = A3\(_{i}\)A2\(_{i}\)-CF * A2\(_{i}\)-CF and identify the i\(^{th}\) image with the maximum A3\(_{i}\)-CF. Bob then attack M with A3\(_{i}\) and if the approximated score matches or is very close to the score given by M, Bob will craft a new set of attack variations similar to that A3\(_{i}\). 
If the given result by M is significantly different from the M' predicted result, Bob will need to put A3\(_{i}\) into M' learned database with values from M, and replace A3\(_{i}\) with a new image.

\item Data poisoning \newline
The ultimate goal is to find attacks that is "mild" enough to be classified around 50\% by M (right at the border of "Attack" and "Normal"). If Bob is able to do that with a right amount of time and occurrences, M may be "tuned" and include Bob's "mild" attacks. After that, Bob will repeat the previous step until the system "favors" Bob enough so he can launch his real attack. This effect is similar to the "boiling frog" effect.
\end{enumerate}
Further details including source codes, images, outputs, etc can be found at Project WolfEye \cite{TamNguyen2017ProjectEye} on Github. Improved version of this demonstration as well as more findings/experiment results will be included in future works.

\section{Recommendations}

\subsection{Invest time on attack/defense model}
From the beginning, ML scientists should pay attention to the ML cyber kill-chain (the attack model) and at least develop a list of recommendations on safe implementation. Recommendations may include but are not limited to the designer's definition of "attack", the meaning of model's accuracy, the side channels, etc. This is essentially important in the context of open-source. As mentioned before, the definition of "attack" should depend on the model's intended purposes rather than just its accuracy. Some ML based solutions were designed to be multi-purposes. Some solutions were originally designed for a specific purpose but were used for other purposes in real-world implementations. Nonetheless, the designers should clearly communicate the intended purposes of the works they are proposing and ways to protect them. For example, most recent works in ML based security for SDNs have accuracy rates of 98\% (Table \ref{table:MLsolutions}) but the meaning of 0.2\% increase in false negatives may differ greatly from one to another. Based on the attack model, the ML designers may also provide a default protection model, explaining how the structure of their designs fit into the protection model, what may be done to harden their works, what are the security trade-offs to be considered, what are the potential side channels in real world deployments, and so on. While research works are not supposed to be commercial ready, basic recommendations on how to protect and harden a ML based solution by its designers are extremely valuable.

\subsection{Design audit-able model}
ML based solutions should generate meaningful logs or even better, having an interface for the model to be audited automatically. Audits may include information on who made what changes, how much the model has drifted after a period of time, the rates of false positives and false negatives, etc. Ideally, the model itself should be able to give indications on whether or not it is under attack and which stage of the kill chain the attackers are at. ML based solution with good audit capability will also help in case the model needs to be rolled back to its earlier versions.

\subsection{Follow secure development processes}
Because ML based security solutions are softwares, the designers should at least follow a secure software development lifecycle \cite{MicrosoftLifecycle}. It involves secure coding practices, static analysis, test cases, attack surface reviews, and so on. Formal verification is absolutely necessary and should be done to the largest extend possible considering there are huge challenges in performing formal verifications on systems like the artificial neural network. Side channels should be limited and there are mechanisms to protect the privacy of the model and its data. Finally, datasets used for training should be as organic as possible.

\subsection{Design an operational cost model}
Cost is another factor as important as accuracy. For the same purpose, a leaner ML algorithm will usually cost less than a complicated one but there may be cases where it is justifiable to have a complex ML model or even a group of different ML models working together. The designers should at least provide a cost model to make practical sense out of their design decisions. A well designed cost model will help with evaluating the cost of false negatives - a very important metric in ML based solutions for cyber security. The paper "Machine Learning with Operational Costs" from MIT researchers \cite{Tulabandhula2013MachineCosts} may serve as a good start for further readings into optimizing ML operational costs. 

\section{Conclusion}
The paper addresses the gap between academic researches on ML based solutions and the operational deployments of ML based systems. While research works do not have to meet the strict requirements for a commercial ready product, it is important that solution designers pay attention and establish some initial foundations for the hardening of their works just in case the works are chosen to be implemented in "the wild". It is important to note that even with a well-funded S.O.C, challenges are still coming from all corners within its systems and from outside. Also, let's not forget that attackers are equipped with Machine Learning powers as well, and can build systems to predict the behaviors of the defending models. For those reasons, this paper proposes a high level threat model tailored specifically to ML based solutions - the ML cyber kill chain - together with four specific recommendations: \textit{\\1) Pay attention to threat models while designing ML solutions.\\2) Make the ML model audit-able\\3) Follow a secure development process\\4)Produce an initial operational cost model}\\
It is believed that these new tools will significantly improve the practical properties of ML based solutions. For the near future, an automatic system will be built and be used to evaluate the robustness of well-known ML based, open source cyber security product such as Apache Spot \cite{ApacheOrganizationApacheSpot}. Hopefully, it can be developed into a threat model assessment tool that ML solution designers can use to provide other metrics in addition to the accuracy rate and prove the robustness of their solutions. 


%\section*{Acknowledgment}




\bibliographystyle{IEEEtran}
\bibliography{IEEEabrv,Mendeley.bib}

\end{document}